\documentclass[twocolumn,superscriptaddress]{revtex4-2}
\usepackage{graphicx}
\usepackage{amsmath}

\begin{document} % length limit: 4 pages except "FULL REFERENCES"

\title{Measurement of small reflectivity anisotropy in the nematic state and its non-equilibrium dynamics}

\author{Inho Kwak}
\author{Jaeseok Son}
\author{Bumjoo Lee}
	\affiliation{Center for Correlated Electron Systems (CCES), Institute for Basic Science (IBS), Seoul 08826, Republic of Korea}
	\affiliation{Department of Physics and Astronomy, Seoul National University, Seoul 08826, Republic of Korea}

\author{Thomas Wolf}
	\affiliation{Institute for Solid State Physics (IFP), Karlsruhe Institute of Technology, 76021 Karlsruhe, Germany}

\author{Changhee Sohn}
	\affiliation{Department of Physics, Ulsan National Institute of Science and Technology, Ulsan 44919, Korea}

\author{Tae Won Noh}
		\thanks{Corresponding author}
		\email{twnoh@snu.ac.kr}
	\affiliation{Center for Correlated Electron Systems (CCES), Institute for Basic Science (IBS), Seoul 08826, Republic of Korea}
	\affiliation{Department of Physics and Astronomy, Seoul National University, Seoul 08826, Republic of Korea}

\author{K. W. Kim}
		\thanks{Corresponding author}
		\email{kyungwan.kim@gmail.com}
	\affiliation{Department of Physics, Chungbuk National University, Cheongju, Chungbuk 28644, Republic of Korea}

\date{\today}

%%%%%%%%%%%%%%%%%%%%%%%%%%%%%%%%%%%%%%%%%%%%%%%%%%%%%%%%%%%%%%%%%%

\begin{abstract} % The abstract should be limited to approximately 100 words. 
Electronic nematicity has attracted a great deal of interest in high-$T_c$ superconductivity. However, measurement of the small optical anisotropy is usually hindered by the geometric anisotropy due to the finite angle of incidence. We present experimental method to investigate nematic anisotropy and its non-equilibrium dynamics. Obtained nematic anisotropy of Ba(Fe$_{0.955}$Co$_{0.045}$)$_{2}$As$_{2}$ single crystal clearly feature the broken four-fold symmetry along orientations of Fe-Fe bonding. Numerical simulations demonstrate that our method is highly reliable in conventional experimental condition. Finally, our time-resolved experiment of nematic anisotropy confirms that ultrafast photo-excitation suppresses the nematic order.
\end{abstract}

\maketitle

%%%%%%%%%%%%%%%%%%%%%%%%%%%%%%%%%%%%%%%%%%%%%%%%%%%%%%%%%%%%%%%%%%

Optical anisotropy appears in various materials bearing lattice, electronic, or magnetic anisotropies. Birefringent materials of large optical anisotropy have been extensively investigated and employed in various optical techniques such as second-harmonic-generations, photo-elastic modulations, and liquid-crystal displays. Even small anisotropy can often play an important role for other physical properties of interest \cite{Tros2015,Oppermann2019,Weightman2005}. 

Recently electronic nematicity, breaking rotational symmetry of lattice, has attracted a great deal of attention with regard to the relation with high-$T_{C}$ superconductivity. The nematic phase precedes the superconductivity in numerous correlated electronic systems such as iron-pnictide \cite{Kuo2016}, cuprate \cite{Daou2010,Sato2017}, heavy Fermion materials \cite{Okazaki2011,Riggs2015}, and topological superconductors \cite{Sun2019}. In Fe-based superconductors, the nematic order in under-doped materials is accompanied with a structural phase transition and exhibits clear electronic anisotropy as confirmed in various experiments such as strain-dependent transport measurement \cite{Kuo2016}, angle-resolved photo-emission spectroscopy \cite{Yi2011}, and optical spectroscopy \cite{Nakajima2012}. In optimal- and over-doped materials, the structure transition is fully suppressed. However, the electronic anisotropy still remains strongly fluctuating and a nematic quantum critical point is supposed to exist near the optimal doping, suggesting that the nematic fluctuations may play an important role for the superconductivity \cite{Kuo2016}. As doping further increases, a new type of nematic order appears in heavily hole-doped materials. That is, the orientation of electronic nematicity is rotated by 45 degrees without any signature of corresponding lattice symmetry breaking. It has been suggested that the change of the nemetic orientation is accompanied with crossover of the superconducting gap symmetry from a nodeless one to a nodal one \cite{Liu2019,Ishida2019}. Therefore, investigation of even small electronic nematicity is important to understand the nematic fluctuations and the superconductivity in Fe-based materials.

Study on a small nematic response requires a challenging signal-to-noise ratio for spectroscopic investigations such as angle-resolved photo-emission spectroscopy. Optical polarimetry experiments using photo-elastic modulator (PEM) with a lock-in technique provide notably high signal-to-noise ratio \cite{Weightman2005,Tros2015,Oppermann2019}. However, measurement of a small optical anisotropy is usually obscured by a geometric anisotropy, which differentiates reflection or transmission coefficients for s- and p-polarized electromagnetic waves. Fresnel equations state that the geometric anisotropy could be considerable at a non-zero angle of incidence even in optically isotropic materials. For transparent materials, the transmission measurements at the perfect normal incidence can be free from the geometric effect. However, the normal incidence to avoid the geometric aniostropy is not feasible for the reflection measurement, inevitable for non-transparent materials. As a result, spectroscopic investigations of the anisotropy of non-transparent materials have been limited or requires only specific experimental conditions such as precise sample-rotation alignments \cite{Weightman2005}. 

%%%%%%%%%%%%%%%%%%%%%%%%%%%%%%%%%%%%%%%%%%%%%%%%%%%%%%%%%%

In this study, we present a new experimental method to investigate the reflectivity anisotropy and its non-equilibrium dynamics. Our method does not require any specific experimental conditions such as precise alignment of the sample rotation angle and the principal axes of the anisotropy are easily characterized. This makes our method appropriate for studying electronic nematicity and its fluctuations in correlated electron systems such as cuprate, pnictide, and heavy Fermion superconductors. As a demonstration of our method, we display experimental results on iron pnictide Ba(Fe$_{0.955}$Co$_{0.045}$)$_{2}$As$_{2}$ single crystals grown from self-flux. Our results clearly shows that the obtained reflectivity anisotropy ${\delta}R/R$ is along the Fe-Fe bonding directions regardless of orientation of the plane of incidence, which usually hinders accurate characterization of a small optical anisotropy. Our numerical simulation using WVASE from Woollam Co. confirms that our method is highly reliable even for a small anisotropy as small as ${\delta}R/R {\sim} 10^{-4}$ order at the angle of incidence in the range up to 15$^{\circ}$. Finally, we perform time-resolved reflectivity anisotropy measurements, confirming that ultrafast pumping with 1.55 eV photon suppresses the reflectivity anisotropy originating from the nematic order. 

%%%%%%%%%%%%%%%%%%%%%%%%%%%%%%%%%%%%%%%%%%%%%%%%%%%%%%%%%%

\begin{figure}[t] %htbp
    \centering
	\includegraphics[width = 1\linewidth]{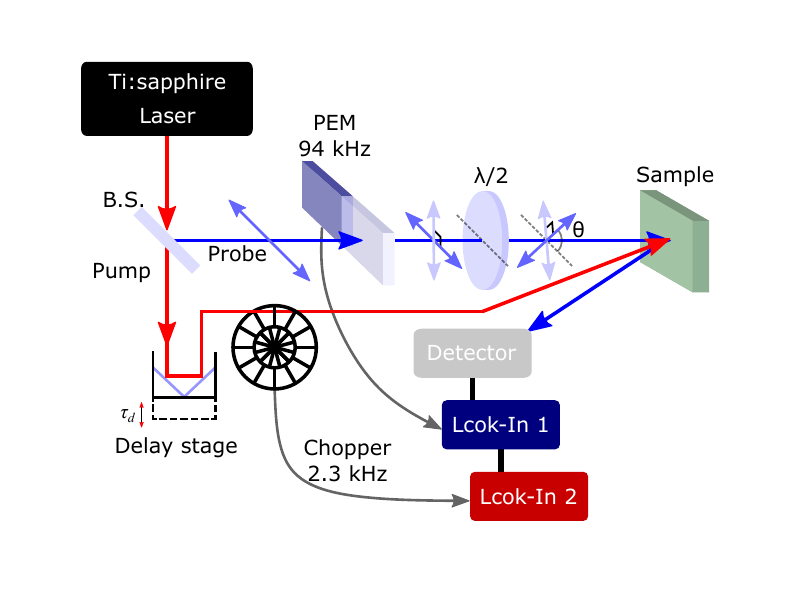}%
	\caption{Schematic diagram of the experimental setup. Ultrafast pulse is split into pump and probe beams after a beam splitter (B.S.). Optic axis of PEM is set to be $45^{\circ}$ with respect to the incoming probe polarization. After the PEM, the probe polarization is varied by rotating half wave-plate.}
    \label{fig:sch}
\end{figure}

We use a standard Ti:sapphire regenerative amplifier which generates pulses of 36 fs pulse duration at full-width-half-maximum (FWHM) and of 1.55 eV photon energy with 250 kHz repetition rate. The generated pulse is split into pump and probe branches, and then they are focused on a sample surface with 100 $\mu$m and 50 $\mu$m in diameter at FWHM for pump and probe, respectively. For polarimetric experiment, we employ a photo-elastic modulator (PEM) providing a sinusoidal modulation of the phase retardation of the probe beam at a frequency $\omega=47$ kHz. The reflectivity anisotropy, that is, the modulation of reflectivity due to the polarization change is obtained as a 2$\omega$ signal through the first lock-in amplifier. Transient change in the reflectivity anisotropy after pumping is measured through the second lock-in amplifier referring to an optical chopper at the pump branch. The pump polarization is set to be parallel to its plane of incidence. Delay time $\tau_{d}$ between pump and probe beams is varied by a delay stage at the pump branch. The experimental setup is schematically displayed in Fig. \ref{fig:sch}. Note that our optical setup is basically same as the one by \textit{Tros et al.} \cite{Tros2015} except that we measure the reflectivity anisotropy of a crystalline sample instead of the transmittance of a liquid sample.

\begin{figure}[t]
    \centering
	\includegraphics[width = 1\linewidth]{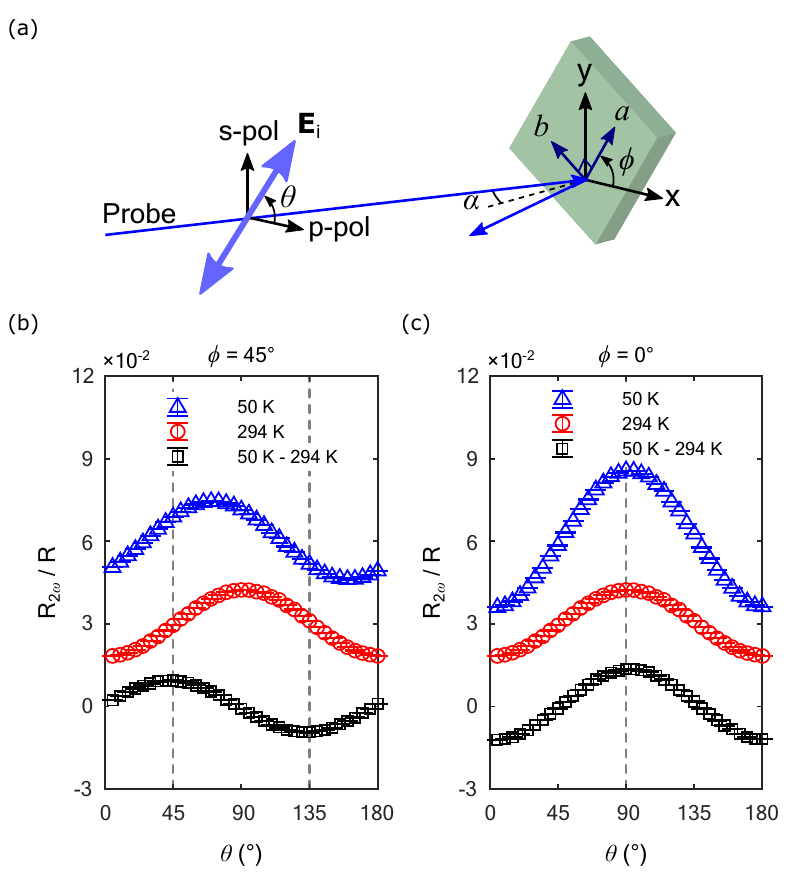}%
	\caption{Experimental geometry of the sample and the probe polarization, and experimental results measured on Ba(Fe$_{0.955}$Co$_{0.045}$)$_2$As$_2$ single crystals. (a) The angles of the probe polarization $\theta$ and of the crystal axis $\phi$ are shown with respect to the plane of incidence. In our experiment, the angle of incidence $\alpha$ is 10$^{\circ}$. The 2$\omega$ lock-in signal $R_{2\omega}/R$ is shown for the cases of (b) $\phi=45^{\circ}$ and (c) $\phi=0^{\circ}$. Data measured in the nematic state at 50 K (blue open triangles) and in the normal state at 294 K (red open circles) are vertically shifted for clarity. Black open squares are obtained by subtracting data at 294 K from data at 50 K.}
	\label{fig:dat}
\end{figure}

The Jones calculus on our experiment shows that the 2$\omega$ lock-in signal $R_{2\omega}$ is a sinusoidal function of the probe polarization angle with respect to the orientation of the crystal anisotropy as follows:
    \begin{align}\label{eq:R2w}
	&{R_{2\omega}(\alpha,\theta,\phi)} = J_2(A) {\times} \nonumber \\
    &\left[
    ({|r_{aa}|^2-|r_{bb}|^2-|r_{ab}|^2+|r_{ba}|^2}) \cos\left(2\theta-2\phi\right)\right. \nonumber \\
    &\qquad \left.- 2\Re\left(r_{aa}^*r_{ab}+r_{ba}^*r_{bb}\right) \sin \left(2\theta-2\phi\right)
    \right],
    \end{align}
{\noindent}where $\alpha$ is the angle of incidence, $\theta$ is the probe polarization angle incident to the sample when it is not modulated by PEM, and $\phi$ is the orientation of a crystal axis as shown in Fig. \ref{fig:dat}(a). $J_2(A)$ is a Bassel function of the first kind and $A$ is the modulated phase angle by PEM. The values of $\theta$ and $\phi$ are indicated with respect to the plane of incidence at the sample surface. $r_{ij}$ is the complex reflectivity coefficient of the sample where $i$ and $j$ indicate polarizations of the incident and the reflected electric fields, respectively. For example, $r_{ab}$ indicate the complex reflectivity coefficient when the incident polarization is parallel to the crystal axis $a$ and the reflected polarization is parallel to the crystal axis $b$. Note that $r_{ij}$ is a function of $\alpha$ and $\phi$, but does not depend on $\theta$. If $\alpha$ is small enough to be considered as a normal incidence case, $|r_{ab}|$ and $|r_{ba}|$ are negligibly small compared to $|r_{aa}|$ and $|r_{bb}|$ \cite{Graham2011}, and Eq. (\ref{eq:R2w}) can be approximated to 
    \begin{equation}\label{eq:R2wa}
    {\frac{R_{2\omega}(\alpha\approx0,\theta,\phi)}{R}} \approx 
    2J_2(A)\frac{|r_{aa}|^2-|r_{bb}|^2}{|r_{aa}|^2+|r_{bb}|^2} \cos\left(2\theta-2\phi\right).
    \end{equation}
In case of half-wave modulation, $A=\pi$ and $2J_2(\pi) = 0.9708$.

Figure \ref{fig:dat} displays experimental data of $R_{2\omega}/R$ measured on Ba(Fe$_{0.955}$Co$_{0.045}$)$_2$As$_2$ single crystals, which shows the nematic phase transition at $T_{\text{nem}}=75$ K. The crystal becomes orthorhombic in the nematic state such that the principal axes of [100]$_{\text{orth}}$ and [010]$_{\text{orth}}$ are parallel to the Fe-Fe bonding directions. We determined the crystallographic orientation by Laue-XRD measurements and prepared two samples on a Cu-plate with different crystal orientations of $\phi=45^{\circ}$ and $0^{\circ}$. Both samples were cleaved before the experiment. In the normal state at 294 K, $R_{2\omega}/R$ (open circles) exhibits the same sinusoidal dependence in both cases of $\phi=45^{\circ}$ in Fig. \ref{fig:dat}(b) and $\phi=0^{\circ}$ in Fig. \ref{fig:dat}(c). We note that this polarization dependence in the sample without in-plane anisotropy originates from the geometric effect in the measurement at non-zero angle of incidence. In the nematic state at 50K, on the other hand, $R_{2\omega}/R$ (open triangles) features a phase-shift from the normal state response for $\phi=45^{\circ}$ in Fig. \ref{fig:dat}(b) while the amplitude is enhanced without a phase-shift for $\phi=0^{\circ}$ in Fig. \ref{fig:dat}(c). This implies that different functions of $\theta$ and $\phi$ due to the nematic order is added up to the normal state signal. These data show that the geometric anisotropy due to the non-zero angle of incidence has a comparable amplitude to the nematic anisotropy.

To remove the geometric anisotropy, we subtract the normal state signal at 294 K (open circles) from the reflectivity anisotropy at 50 K (open triangles):
\begin{equation}\label{eq:dR}
    \frac{{\delta}R_{2\omega}}{R} =
    \left.{\frac{R_{2\omega}(\alpha,\theta,\phi)}{R}}\right|_{T<T_{\text{nem}}} - \left.{\frac{R_{2\omega}(\alpha,\theta,\phi)}{R}}\right|_{294 \text{K}}.
\end{equation}
{\noindent}Obtained nematic anisotropies (open squares) clearly features a cosine function of $(2\theta-2\phi)$, as displayed in Fig. \ref{fig:dat}(b) for $\phi=45^{\circ}$ and Fig. \ref{fig:dat}(c) for $\phi=0^{\circ}$, which agrees with Eq. (\ref{eq:R2wa}).

\begin{figure}[t]
    \centering
	\includegraphics[width = 1\linewidth]{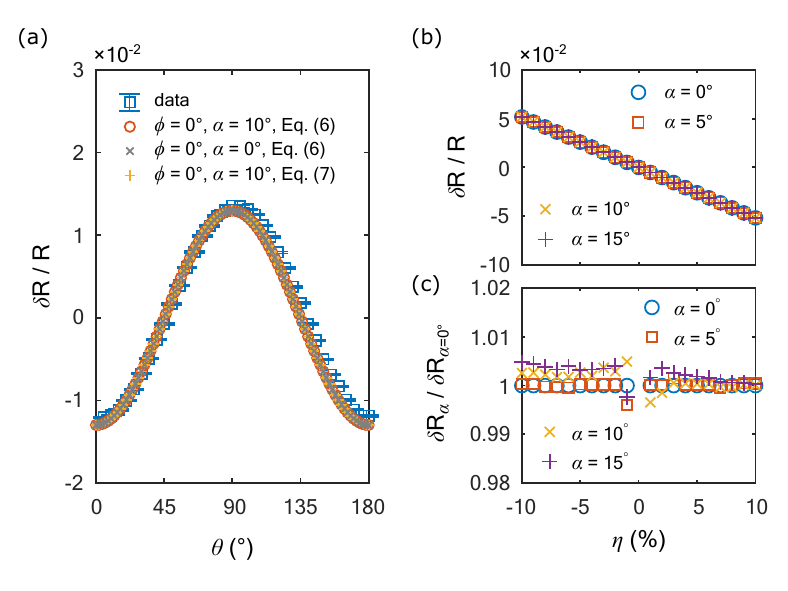}%
	\caption{Simulations of the reflectivity anisotropy. (a) Comparison of the experimentally obtained nematic anisotropy (open square) with simulated ones for $\alpha = 10^{\circ}$ and $\phi=0^{\circ}$ by Eq. (\ref{eq:dRsm}) and Eq. (\ref{eq:dRr}). All of them agree with the the exact case of $\alpha=0^{\circ}$ at $\phi=0^{\circ}$ (symbol $\times$).  (b) Nematic anisotropy simulated for various values of $\alpha$ by Eq. (\ref{eq:dRsm}) and (c) the normalized representation by the normal incidence case are shown as a function of relative change of refractive index ${\eta}$. Refractive index $n_a$ and $n_b$ for the nematic state is varied as described in the text.
	}
	\label{fig:sim}
\end{figure}

Numerical simulation of the reflectivity anisotropy confirms that the experimental approach according to Eq. (\ref{eq:dR}) can successfully represent the nematic anisotropy as displayed in Fig. \ref{fig:sim}(a). For the simulation on the normal state, we employed a virtual refractive index of $n_{ab}=2.865+2.53i$, and $n_c=1.62+2.05i$ at the photon energy of 1.55 eV. The value of $n_{ab}$ is from the experimental results on BaFe$_2$As$_2$ single crystal measured at room temperature by spectroscopic Ellipsometer, and the value of $n_c$ is extracted from published data \cite{Moon2013}. For the nematic state, we increased (decreased) the real part and decreased (increased) the imaginary part of $n_a$ ($n_{b}$) by a relative ratio $\eta$ such that ${\delta}n=n_{a}-n_{b} \propto 2{\eta} {\times} n_{ab}^{*}$, which is consistent to the change observed in the nematic state \cite{Moon2013}. Reflectivity of s-polarization $R_s$ and p-polarization $R_p$ for $\phi=0^{\circ}$ were calculated using WVASE (Woollam Co.). Reflectivity and its anisotropy at arbitrary polarization were calculated as below:
\begin{equation}\label{eq:Rth}
    R(\alpha,\theta)=R_p(\alpha)\cos^2 \theta + R_s(\alpha)\sin^2 \theta,
\end{equation}
\begin{equation}\label{eq:Rsm}
    \frac{R_{\text{sim}}(\alpha,\theta)}{R}=
    \frac{R(\alpha,\theta)-R(\alpha,\theta+90^{\circ})}
         {R(\alpha,\theta)+R(\alpha,\theta+90^{\circ})},
\end{equation}
\begin{equation}\label{eq:dRsm}
    \frac{{\delta}R_{\text{sim}}}{R}=
    \left.\frac{R_{\text{sim}}(\alpha,\theta)}{R}\right|_{n_a{\neq}n_b} - \left.\frac{R_{\text{sim}}(\alpha,\theta)}{R}\right|_{n_a=n_b}.
\end{equation}
{\noindent}As displayed in Fig. \ref{fig:sim}(a), our experimental result (open square) of $\phi=0^{\circ}$ by Eq. (\ref{eq:dR}) is well reproduced by the numerical simulation (open circle) of Eq. (\ref{eq:dRsm}) with $\eta=5$ $\%$ variation of the refractive index for the nematic state. Furthermore, they match well with the true reflectivity anisotropy of the material for the case of the normal incidence (symbol $\times$), where the geometric anisotropy is absent.

\begin{figure}[t]
    \centering
	\includegraphics[width = 1\linewidth]{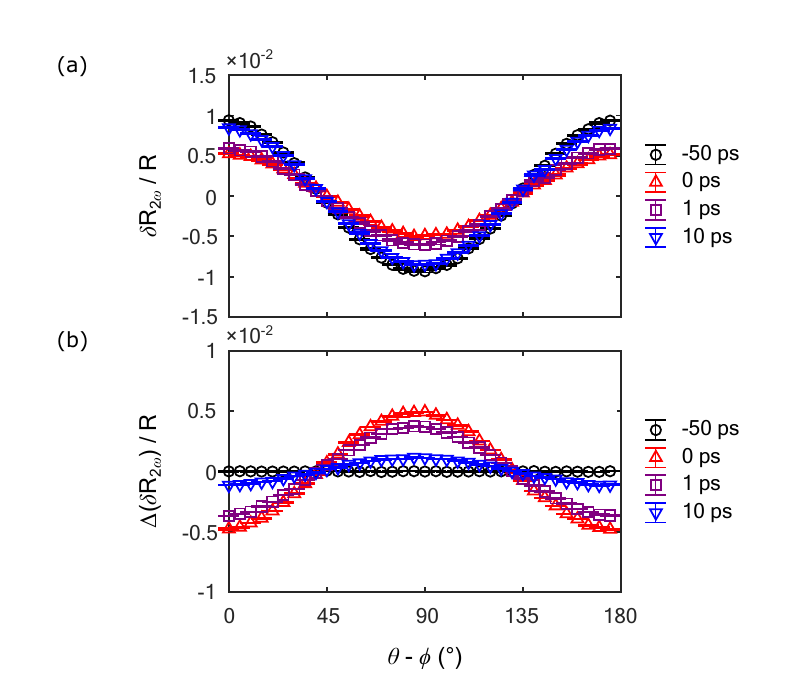}%
	\caption{Pump-probe results of the reflectivity anisotropy. (a) Nematic anisotropy ${\delta}R/R$ of Ba(Fe$_{0.955}$Co$_{0.045}$)$_2$As$_2$ single crystal at 50 K obtained by Eq. (\ref{eq:dR}) and (b) its transient change ${\Delta}({\delta}R)/R$ measured at various delay times ${\tau}_{d}$. The pump fluence is fixed at 31 ${\mu}$J/cm$^2$.}
	\label{fig:und}
\end{figure}

To demonstrate the reliability of Eq. (\ref{eq:dR}), we performed the simulation for $\alpha \leq 15^{\circ}$ and for $\eta$ up to $\pm 10 {\%}$. As displayed in Fig. \ref{fig:sim}(b), ${\delta}R_{\text{sim}}/R$ is proportional to $\eta$. The angle of incidence $\alpha$ does not affect the result within the order of $10^{-4}$ of the absolute reflectivity anisotropy ratio as displayed in Figs. \ref{fig:sim}(b) and (c).

We note that our method can be extended to materials of small nematicity even when the high temperature isotropic phase is not accessible. In a nematic state, nematic twin domains are easily found in a single piece of a sample. Because the geometric anisotropy is expected to be similar in the two twin domains, we can eliminate the geometric effect from the difference of the reflectivity measured on the two different domains. For the given case of the crystal orientation of $\phi$, the twin domain should have $\phi^{\prime}=\phi+90^{\circ}$. Therefore, the reflectivity anisotropy can be obtained as follows:
\begin{equation}\label{eq:dRr}
    \frac{{\delta}R(\theta,\phi)}{R} = \frac{
    R_{2\omega}(\theta,\phi) - R_{2\omega}(\theta,\phi+90^{\circ})}
    {2R}.
\end{equation}
{\noindent}One can measure on the twin domains by simply translating the sample surface, which should be appropriate in most of samples in a nematic state. Such a measurement could be performed also on a single domain sample as well by rotating the sample if the crystal angle could be precisely controlled. Figure \ref{fig:sim}(a) shows that the result from this method also agrees nicely with the true reflectivity anisotropy for the normal incidence case.

Finally we measure transient reflectivity anisotropy ${\Delta}({\delta}R)/R$ of Ba(Fe$_{0.955}$Co$_{0.045}$)$_2$As$_2$ single crystal at 50 K upon pumping with 800 nm photons. Our results clearly show that the reflectivity anisotropy due to the nematic order is suppressed by ultrafast photo-excitation as displayed in Fig. \ref{fig:und}(a). The nematic responses of Fe-based superconductors have been observed in pump-probe experiments even in the tetragonal structure. However, because the pumping with linearly polarized photons inherently possesses an anisotropic nature, the observed nematicity in the pump-probe experiment can be questioned whether it is the intrinsic nematicity. We found that the change of the reflectivity anisotropy upon pumping does not depend on the polarization of the pump pulses. Our simultaneous observation of ${\delta}R/R$ and ${\Delta}({\delta}R)/R$ demonstrates that the nematic order is suppressed by the photo-excitation without an pump-induced anisotropy and, therefore, assures that previous time-resolved studies on the nematicity indeed reflects the intrinsic nematic properties of the iron pnictides \cite{Patz2014,Shimojima2019}.

In summary, we present experimental method to study the optical anisotropy by reflection experiment. As a demonstration, we measure the nematic anisotropy of Ba(Fe$_{0.955}$Co$_{0.045}$)$_2$As$_2$ single crystal. Numerical simulations confirm that the experimentally observed data are highly reliable. Successive time-resolved experiment on the nematic anisotropy assures that the ultrafast photo-excitation simply suppresses the nematic anisotropy of the iron pnictides.

\medskip
\noindent\textbf{Funding.}
Institute for Basic Science (IBS) in Korea (IBS-R009-D1); National Research Foundation of Korea (NRF) (NRF-2017M3D1A1040828, 2017R1A4A1015323, and 2019R1F1A1062847).
%Funding information should be listed in a separate block preceding any acknowledgments. List just the funding agencies and any associated grants or project numbers, as shown in the example below:
%National Science Foundation (NSF) (1263236, 0968895, 1102301); The 863 Program (2013AA014402).
%The acknowledgments may contain any information that is not related to funding:

\medskip
\noindent\textbf{Acknowledgment.}
The authors thank J. S. Lee (GIST) for valuable discussions.

\medskip
\noindent\textbf{Disclosures.} The authors declare no conflicts of interest.

% Bibliography
\bibliography{OPP_PEM_arXiv.bib}

\end{document}